 \documentclass[aps, twocolumn, amsmath]{revtex4}
\usepackage{dcolumn}
\usepackage{bm}
\usepackage{graphicx}
\usepackage{color}
\usepackage{subfigure}
\usepackage{hyperref}
\usepackage{latexsym}
\usepackage{amsthm}
\usepackage{amssymb}
\DeclareGraphicsExtensions{.jpg,.pdf, .mps, .png, .eps, .ps, .EPS,.gif}
\begin{document}
\def\be{\begin{equation}}
\def\ee{\end{equation}}

\def\bc{\begin{center}}
\def\ec{\end{center}}
\def\bea{\begin{eqnarray}}
\def\eea{\end{eqnarray}}
\newcommand{\avg}[1]{\langle{#1}\rangle}
\newcommand{\Avg}[1]{\left\langle{#1}\right\rangle}

\def\ie{\textit{i.e.}}
\def\etal{\textit{et al.}}
\def\m{\vec{m}}
\def\G{\mathcal{G}}

\title{Rare events and discontinuous percolation transitions }

\author{ Ginestra Bianconi}
\affiliation{School of Mathematical Sciences, Queen Mary University of London, London, E1 4NS, United Kingdom}

\begin{abstract}
Percolation theory  characterizing the robustness of a  network  has applications ranging from biology, to epidemic spreading,  and complex infrastructures. Percolation theory, however, only concern the typical response of an infinite  network to random damage of its nodes while in real  finite networks, fluctuations  are observable. Consequently for finite networks there is an urgent need  to evaluate the  risk of collapse in response to rare configurations of the initial damage.  Here we  build a large deviation theory of percolation characterizing  the response of a sparse network to  rare events. This  general theory includes the second order phase transition observed  typically for random configurations of the initial damage but reveals also discontinuous transitions corresponding to  rare configurations of the initial damage  for which  the size of the giant component is suppressed.
\end{abstract}

\maketitle

\section{Introduction}
Percolation theory  \cite{crit,Kahng_review,Newman_old1,Laszlo_robustness,Cohen1,Cohen2} plays a pivotal role in characterizing the robustness of a network as it sheds light on the fundamental structural properties  that determine its response when a fraction of  nodes is initially damaged. Therefore  percolation theory is a fundamental critical phenomena that  permeates statistical mechanics as well as  network science \cite{NS,Newman_book,Havlin_book} having profound implications in different contexts ranging from   ecological networks to  infrastructures.

Despite the fact that the percolation transition is second order, cascade of failure events that abruptly dismantle a network are actually occurring in real systems, with major examples ranging from large electric blackouts to the sudden collapse of ecological systems.
In order to explain how abrupt phase transitions could result from  percolation,  recently generalized percolation  problems including percolation in interdependent multilayer networks \cite{Havlin1,Doro_multiplex,Son,Redundant,Cellai2013,Radicchi,Havlin2}, and explosive percolation \cite{Explosive,Doro_explosive,Riordan_explosive,Grassberger_explosive} that retards the percolation transition, have been proposed. It has been shown that in interdependent multilayer networks  discontinuous phase transitions are the rule \cite{Havlin1,Doro_multiplex,Son,Redundant,Cellai2013,Radicchi,Havlin2}. For explosive percolation it has been proved  that  the original Achiloptas process \cite{Explosive,Doro_explosive,Riordan_explosive,Grassberger_explosive}  yields  a steep but continuous transition despite some of its modifications are  currently believed to yield genuinely discontinuous  transitions \cite{Herrmann,Kahng,Souza}. 
It is to note that this interest on discontinous percolation transitions has triggered further research in the statistical mechanics of networks. In fact discontinuous phase transitions have been observed also  in explosive synchronization of single and multilayer networks \cite{Arenas,Vito,Boccaletti} in the wider context of the Kuramoto dynamics previously believed to yield exclusively  second order transitions.

Simple node percolation \cite{Newman_old1,Laszlo_robustness,Cohen1,Cohen2} has been one of the most investigated critical phenomena on networks. It determines the response of the network to a random initial damage. Since belonging to the giant component is often  considered a pre-requisit for the node to be functional, all the nodes  that are not any more in the giant component are assumed to fail as a consequence of the initial damage. Therefore characterizing the  percolation  transition on a single network is widely considered as a simple yet powerful way  to evaluate the robustness of a network. Despite recently some  attention has been drawn to the characterization of extremal configurations of the initial damage that dismantle most efficiently complex networks \cite{Makse,Dismantling,Fluct1}, the vast majority of the scientific research  concerns so far the typical scenario characterized by the well known continous second order phase transition \cite{Newman_old1,Laszlo_robustness,Cohen1,Cohen2}.

In infinite networks percolation is known to be self-averaging, i.e. fluctuations from the typical behavior are vanishing. However in finite real networks rare events are observable and it is of fundamental importance to have a  complete theoretical framework for characterizing the response of the network also to rare configurations of the initial damage. Here we address this problem by  investigating  the large deviations \cite{Touchette} of  percolation on sparse networks. We show that percolation theory on single networks includes both continuous and discontinuous phase transitions as long as we consider also rare events.  The entire phase diagram of percolation is  uncovered using naturally defined   thermodynamic quantities  including the free energy, the entropy and the specific heat of percolation. The continuous phase transition dominating the typical behavior is derived in the context of this more general theoretical approach. Additionally we  observe that rare configurations  of the damage yield   discontinuous phase transitions whereas  the imposed bias on the configurations of the initial damage tends to suppress the  size of the giant component.   These results shed new light on possible mechanisms yielding abrupt phase transitions \cite{ecosystems} and might play a crucial role for determining early warning signals of these transitions. Using the theory of large deviations we show   that the observed   discontinuous phase transitions  are caused by the fact that particularly damaging initial damage configurations can be observable in finite networks.

It is well known that the percolation transition can be studied by investigating the Potts model in the limit in which the spins can be in $q\to 1$ states \cite{FK,Wu}. Interestingly the Potts formalism has been also used to explore the large deviation of the number of clusters in  random and complex networks \cite{Monasson, Bradde}. Our approach is rather distinct from these previous studies  because we are not concerned with the probability of observing a certain number of clusters, but instead we focus on the probability of the initial damage configurations that yield a  given size of the giant component. We note here that while   the number of clusters does not determine the properties of the percolation transition, the size of the giant component is nothing else that  the order parameter of percolation and therefore  it is the key quantity determining the transition.

Our approach, based on a locally tree-like approximation, uses a message passing algorithm, specifically Belief  Propagation   \cite{Mezard,Weigt,Yedida,Semerjian}.
Message passing algorithms are becoming increasingly relevant in the context of complex networks and have been recently widely used for percolation \cite{Cellai2013,Radicchi,Lenka}, epidemic spreading \cite{Epidemics_Luca,Saad,Gleeson_MP} and  network control \cite{Control,Bianconi_control}.
The proposed   Belief Propagation algorithm  reveals the large deviation of percolation and characterizes its phase diagram on single network realizations including real network datasets and single instances of random network ensembles. Here we apply this theoretical framework both to real datasets of foodwebs and to uncorrelated network ensembles.

The paper is organized as follows: in Sec. II we describe the large deviation approach to percolation, in Sec. III we provide the detailed Belief Propagation equations that solve the large deviation properties of percolation on single networks, in Sec. IV we characterize the equations determining the large deviation of percolation in network ensembles. In Sec. V we provide analytical evidence of the discontinuous phase transition observed for regular networks as soon as the giant component is suppressed and we study the large deviation properties of percolation on Poisson networks and real foodwebs using the BP algorithm. Finally in Sec VI we provide the conclusions.

\section{The large deviation approach to percolation}
\label{secBP}
\subsection{Message passing algorithm on single realization of damage}
\begin{figure*}[!htb]
    \includegraphics[width=1.90\columnwidth]{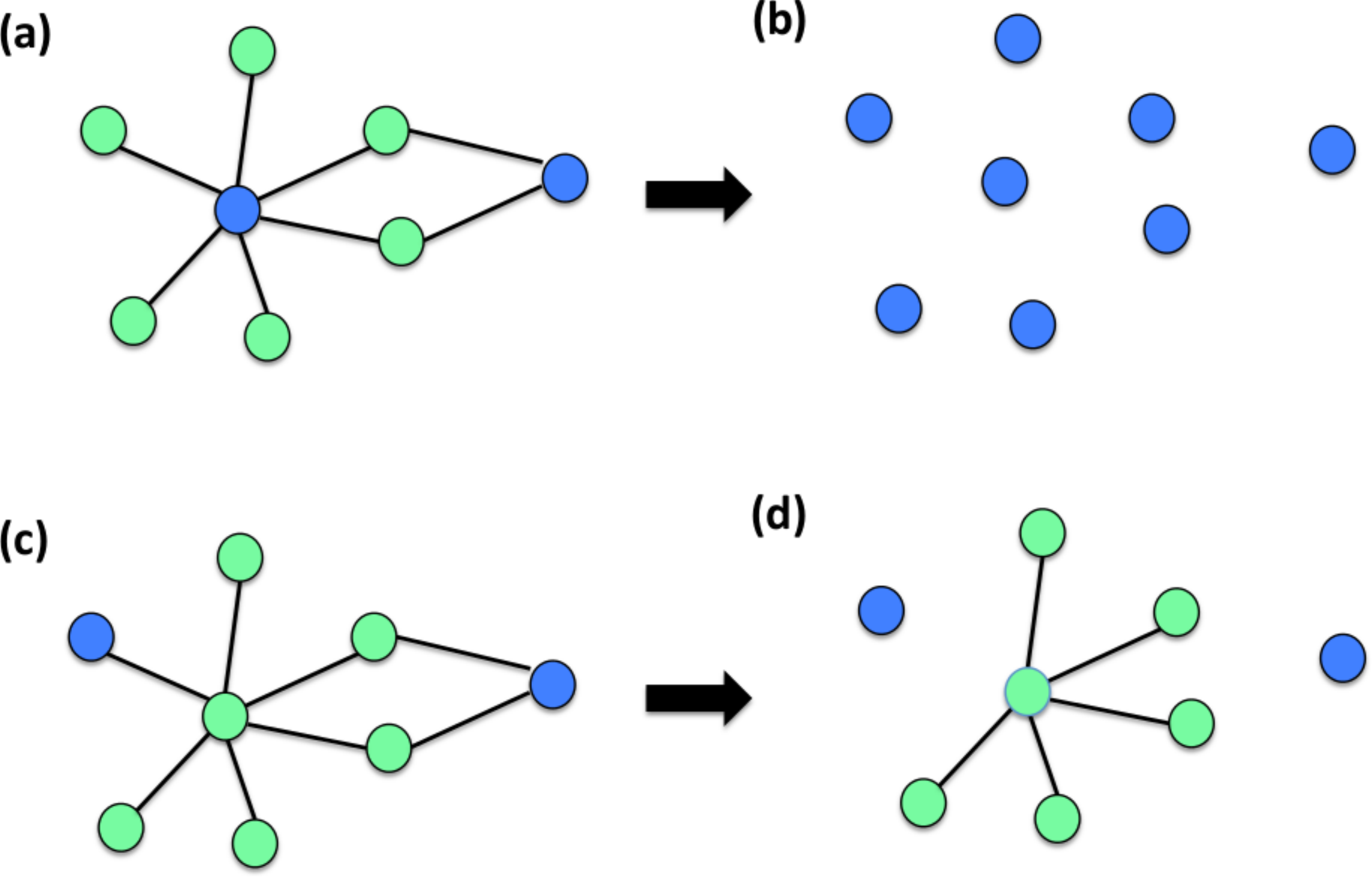}
	\caption{ Different realizations of the initial damage are here shown to be  more or less damaging for a network. Panel (a) shows an initial damage of a connected network  affecting exclusively two  out of the $N=8$ nodes of the network (blue nodes indicate damaged nodes green nodes indicate not damaged nodes). Panel (b) shows that this initial damage is very disruptive for the network  and results in giant component of size $R=1$. Panel (c) shows another initial damage configuration of the same network which affects only two nodes of network. In this case panel (d) shows that the effect of the damage are reduced and most of the network remains connected resulting in a giant component $R=6$. }
	\label{fig:sketch}
\end{figure*}

Let us consider a given locally tree-like network of $N$ nodes  
where each node $i=1,2,\ldots, N$ is either damaged ($x_i=0$) or not ($x_i=1$).
In this case it is well known that the following message passing algorithm is able to determine whether a node belongs ($\rho_i=1$) or not ($\rho_i=0$) to the giant component.
Specifically the message passing algorithm consists on a set or recursive equations written for the messages $\sigma_{i\to j}$ that each node $i$ send to a neighbour node $j$ of the network. (Note that of each interaction between node $i$ and node $j$ there are two distinct messages  $\sigma_{i\to j}$ and $\sigma_{j\to i}$).
The message passing equations read,
\bea
\sigma_{i\to j}=x_i\left[1-\prod_{\ell\in N(i)\setminus j}(1-\sigma_{\ell\to i})\right],
\label{mess}
\eea
where $N(i)$ indicates the set of neighbours of node $i$.
The messages $\sigma_{i\to j}$ that  $\rho_i$ which is given by 
\bea
\rho_i=x_i\left[1-\prod_{j\in N(i)}(1-\sigma_{j\to i})\right].
\eea
Finally the size of the giant component of the network ${\mathcal R}$, resulting after the inflicted initial damage $\{x_i\}_{i=1,2,\ldots, N}$ is given by 
\bea
{\mathcal R}=\sum_{i=1}^N\rho_i.
\eea
Therefore different realizations of the initial damage can yield, in general, giant components of different sizes (see schematic discussion in Figure \ref{fig:sketch}).

In the following we will indicate with $\bm{\sigma}$ the set of all the messages and with $\bm{\sigma}_i$ the set of all the messages starting or ending at node $i$, i.e.
\bea
\bm{\sigma}&=&\{\sigma_{i\to j}\}_{i\in\{1,2,\ldots, N\}; j\in N(i)},\nonumber \\
\bm{\sigma}_i&=&\{\sigma_{i\to j},\sigma_{j\to i}\}_{ j\in N(i)}.
\eea
Additionally we will indicate with $\bm{x}$  the configuration of the initial damage, i.e.
\bea
\bm{x}=\{x_i\}_{i\in \{1,2,\ldots, N\}}.
\eea

\subsection{Random realization of the damage and typical behaviour}

Here we are concerned with realizations of the initial damage $\bm{x}$ where each node is damaged with probability $1-p$, i.e. each configuration $\bm{x}$ is drawn from a distribution
\bea
\tilde{P}(\bm{x})=\prod_{i=1}^N p^{x_i}(1-p)^{1-x_i}.
\label{random}
\eea
Usually, in order to predict the expected size of the giant component $\hat{R}$ given by 
\bea
\hat{R}=\sum_{\bm{x}}\tilde{P}(\bm{x}) {\mathcal R}, 
\eea
the original message passing algorithm is averaged over the distribution $\tilde{P}(\bm{x})$.
Given the locally tree-like structure of the network this procedure generates a novel message passing algorithm determined by the set of messages 
\bea
\hat{\sigma}_{i\to j}=\sum_{\bm{x}}\tilde{P}(\bm{x}) \sigma_{i\to j},
\eea
 satisfying 
\bea
\hat{\sigma}_{i\to j}=p\left[1-\prod_{\ell\in N(i)\setminus j}(1-\hat{\sigma}_{\ell\to i})\right].
\label{mess_av}
\eea 
These messages determine the probability 
\bea
\hat{\rho}_i=\sum_{\bm{x}}\tilde{P}(\bm{x}) \rho_i
\eea 
that node $i$ is in the giant component,  which is given by 
\bea
\hat{\rho_i}=p\left[1-\prod_{\ell\in N(i)}(1-\hat{\sigma}_{\ell\to i})\right].
\eea
Finally the expected size of the giant component $\hat{R}$ is given by 
\bea
\hat{R}=\sum_{i=1}^N \hat{\rho}_i.
\eea

\subsection{Large deviations of percolation}

Here  we are interested in going beyond the typical scenario by characterizing the probability $\pi(R)$ that a given configuration of the initial damage yields  a giant component of size $R$, i.e.
\bea
\pi(R)=\sum_{\bm{x}}\tilde{P}(\bm{x}) \delta({\mathcal R},R),
\label{pi_def}
\eea
where $\delta(m,n)$ is the Kronecker delta.
For any given value of $p$, and for large network sizes $N\gg 1$ the probability $\pi(R)$ will follow the large deviation scaling \cite{Touchette}
\bea
\pi({R})\sim e^{-NI({R})},
\eea
where $I(R)\geq 0$ is called the {\em rate function}. This expression indicates that for any given value of $p$ the deviations from the most likely size of the giant component are exponentially suppressed. Additionally this expression implies that on an  infinite network the percolation transition is self-averaging and that all networks with yield almost surely the same giant component $R=\hat{R}$ for which $I(R)$ takes its minimum value $I(\hat{R})$.
In order to find $I(R)$ let us introduce the partition function $Z=Z(\omega)$ 
\bea
Z&=&\sum_{\bm{x}}\tilde{P}(\bm{x}) e^{-\omega {\mathcal R}}.
\eea
Using the definition of $\pi(R)$ given by Eq. $(\ref{pi_def})$ it can be easily shown that 
$Z$ is the generating function of $\pi(R)$ as $Z$  can  be written as 
\bea
Z=\sum_{R}\pi({R})e^{-\omega {R}}
\label{Zdef}
\eea
By indicating with $F$ the corresponding free-energy and with $f$ the free energy density given by 
\bea
\omega F=\omega Nf=-\log(Z).
\eea
it is immediate to show that $\omega f(\omega)$ a is the  Legendre-Fenchel tranforms of  the rate function $I(R)$ \cite{Touchette}.
In particular we have that $\omega f(\omega)$ can be expressed as 
\bea
\omega f(\omega)&=&\inf_{R}\left[I(R)+\omega \frac{R}{N}\right].
\eea
Additionally as long $\omega f(\omega)$ is differentiable,  the Legendre-Fenchel transform of   $\omega f(\omega)$ fully determines $I(R)$, given by the convex function
\bea
I(R)=\sup_{\omega} \left[\omega f(\omega)-\omega \frac{R}{N}\right]
\eea
 
Therefore as long as $\omega f(\omega)$ is differentiable, by studying the free energy $\omega f(\omega)$ of the percolation problem the large deviation of the size of the giant component can be fully established and the rate function $I(R)$ is convex. However when $I(R)$ is non-convex  $\omega f(\omega)$ is not differentiable and the Legendre-Fenchel transform of $\omega f(\omega)$ only provides the convex envelop of $I(R)$ \cite{Touchette}.

\subsection{The Gibbs measure over messages}

In order to study the partition function $Z$, we make a change of variables and instead of considering a Gibbs measure over configurations of the initial damage we  consider the Gibbs measure $P(\bm{\sigma})$ over the set $\bm{\sigma}$ of all messages. The probability $P(\bm{\sigma})$  allows us to determine the most likely distribution of the messages corresponding to a given  size of the giant component ${\mathcal R}$. The large deviations properties of percolation are studied by  introducing a  Lagrangian multiplier $\omega$  modulating the average size of the giant component ${\mathcal R}$.Therefore the  Gibbs measure $P(\bm{\sigma})$  is given by 
\bea
P({\bm{\sigma}})=\frac{1}{Z}\sum_{{\bm{x}}}e^{-\omega {\mathcal R}}\tilde{P}({\bf x})\chi(\bm{\sigma},\bm{x}),
\label{Pss}
\eea
where the function $\chi(\bm{\sigma},\bm{x})$ enforces the message passing Eqs. $(\ref{mess})$, i.e.
\bea
\chi(\bm{\sigma},\bm{x})=\prod_{i=1}^N\prod_{j\in N(i)}\delta\left(\sigma_{i\to j},x_i\left[1-\prod_{\ell\in N(i)}(1-\sigma_{\ell\to i})\right]\right).\nonumber
\eea
Here $Z$ is the partition function of the problem and it can be easily shown that it reduces to $Z$ defined in Eq. (\ref{Zdef}), i.e.
\bea
Z=\sum_{\bm{\sigma}}\sum_{{\bm{x}}}e^{-\omega {\mathcal R}}\tilde{P}({\bm{x}})\chi({\bm{x}},\bm{\sigma})=\sum_{R}\pi(R)e^{-\omega {R}}.
\eea 
The role of $\omega$ in determining the Gibbs measure $P(\bm{\sigma})$ is equivalent to the one of temperature in a canonical ensemble. Since for each node only two options are possible: either a node belongs  $(\rho_i=1)$ or do not ($\rho_i=0$) belong to the giant component, this problem can be interpreted as a statistical mechanics problem of a  two level systems. Therefore it is possible to investigate the role of both positive and negative values of $\omega$.

For $\omega<0$, the Gibbs measure weights more the {\em buffering } configurations of the  initial damage resulting in a giant component larger than the typical one.
On the contrary for $\omega>0$ the Gibbs measure weights more the {\em aggravating} configurations of the initial damage resulting in a giant component smaller than the typical one.
For $\omega=0$ we recover the typical scenario.

From Eq. $(\ref{Pss})$ it  follows that $P(\bm{\sigma})$ can be expressed as 
\bea
P(\bm{\sigma})=\frac{1}{Z}\prod_{i=1}^N \psi_i(\bm{\sigma}_i,\omega),
\label{GM}
\eea
where the set of constraints $\psi_i(\bm{\sigma}_i,\omega)$ for $i=1, 2, \ldots, N$ defined over all the messages $\bm{\sigma}_i$ starting or ending to node $i$ read
\bea
\psi_i(\bm{\sigma}_i)&=&\left[(1-p)\prod_{j\in N(i)}\delta(\sigma_{i\to j},0)\right.\nonumber \\
&&\hspace*{-30mm}+\left.p e^{-\omega \hat{\rho_i}}\prod_{j\in N(i)}\delta\left(\sigma_{i\to j},1-\prod_{\ell\in N(i)\setminus j}(1-\sigma_{\ell \to i})\right)\right],
\eea
where $\delta(m,n)$ indicates the Kronecker delta and $\hat{\rho}_i$ is given by 
\bea
\hat{\rho_i}=\left[1-\prod_{j\in N(i)}(1-\sigma_{j\to i})\right].
\eea
Given Eq. $(\ref{GM})$ it follows that the partition function  $Z$ can be also written as  
\bea
Z=\sum_{\bm{\sigma}}\prod_{i=1}^N \psi_i(\bm{\sigma}_i,\omega).
\eea 

From this theoretical framework it is possible to derive naturally the following thermodynamic quantities for percolation:  energy $R$, free energy $F$, entropy $S$ and specific heat $C$.
Specifically the  energy $R$ is the average size of the giant component of the network, the free energy $F$ is proportional to the logarithm of the partition function $Z$ with $\omega F(\omega)/N$ indicating  the Legendre-Fenchel transform of the rate function $I(R)$,  the entropy $S$ determines the logarithm of the typical number of message configurations that yield a given size of the giant component $R$ and  the specific heat $C$ is proportional  to the variance  of the giant component  for  given values of $p$ and $\omega$  (see Table $\ref{ther_quantities}$). 

\begin{table*}[!htb]
\begin{tabular}{|l|l|}
\hline
\hline
Thermodynamic quantities & Mathematical relations\\
\hline
{Energy}  $R$ & $R=-\frac{\partial \ln Z}{\partial\omega}$\\
{Free energy} $F$ & $\omega F=-\ln Z$\\
{Entropy} $S$ & $S=-\sum_{\bm{\sigma}}  P(\bm{\sigma})\ln P(\bm{\sigma})$ \\
{Specific heat} $C$& $C=\omega^2\frac{\partial^2 \ln Z}{\partial\omega^2}$ \\
\hline
\end{tabular}
\caption{The thermodynamic quantities of percolation (energy $R$, free energy $F$, entropy $S$ and specific heat $C$) are listed together with  their mathematical expression in terms of the probability $P(\bm{\sigma})$ and its associated partition function $Z$.}
\label{ther_quantities}
\end{table*}

The Gibbs measure and the corresponding thermodynamic quantities can be calculated in the locally tree-like approximation using Belief Propagation (BP) for any given locally tree-like network, representing either a real network  dataset or a single instance of a random network model. Moreover the BP equations can be also averaged  over network ensembles with degree distribution $P(k)$ characterizing the nature of the phase transition (see next two sections).

\section{Large deviation theory of percolation on single networks}
\subsubsection{The Belief Propagation equations }
The Gibbs distribution $P(\bm{\sigma})$   can be expressed explicitly on a locally tree-like network using the  Belief Propagation  (BP) method \cite{Mezard,Yedida, Weigt,Semerjian} by finding the messages 
$\hat{P}_{i\to j}(\sigma_{i \to j},\sigma_{j\to i})$ that each  node $i$  sends to the generic neighbour node  $j$.
These message satisfy the following recursive BP equations
\bea
\hspace*{-7mm}\hat{P}_{i\to j}(\sigma_{i \to j},\sigma_{j\to i})=\frac{1}{{\mathcal C}_{i\to j}}\sum_{\bm{\sigma}_i} \psi_i(\bm{\sigma}_i) \prod_{\ell\in N(i)\setminus j}\hat{P}_{\ell\to i}(\sigma_{\ell\to i},\sigma_{i\to \ell}),\nonumber
\label{BP}
\eea
where ${\mathcal C}_{i\to j}$ are normalization constants enforcing the normalization condition 
\bea
\sum_{\sigma_{i\to j}=0,1}\sum_{\sigma_{j\to i}=0,1}\hat{P}_{i\to j}(\sigma_{i\to j},\sigma_{j\to i})=1.
\eea
In the Bethe approximation, valid on locally tree-like networks the probability distribution $P(\bm{\sigma})$ is given by 
\bea
P(\bm{\sigma})&=&\prod_{i=1}^N {\mathcal P}_i(\bm{\sigma}_i)\left(\prod_{<i,j>}{\mathcal P}_{ij}(\sigma_{i \to j},\sigma_{j \to i})\right)^{-1}
\label{uno}
\eea
where  ${\mathcal P}_{i}(\bm{\sigma}_i)$ and ${\mathcal P}_{ij}(\sigma_{i \to j},\sigma_{j \to i})$ indicate the marginal distribution of nodes and links and are given  by 
 \bea
{\mathcal P}_{ij}(\sigma_{i \to j},\sigma_{j \to i})&=&\frac{1}{\mathcal{C}_{ij}}\hat{P}_{i\to j}(\sigma_{i \to j},\sigma_{j \to i})\hat{P}_{j\to i}(\sigma_{j \to i},\sigma_{i \to j}),\nonumber \\
 {\mathcal P}_{i}(\bm{\sigma}_i)&=&\frac{1}{\mathcal{C}_{i}} \psi_i(\bm{\sigma}_i)\prod_{j\in N(i)}\hat{P}_{j\to i}(\sigma_{j \to i},\sigma_{i \to j}),
\label{marginals}
\eea
with ${\mathcal C}_i$ and ${\mathcal C}_{ij}$ indicating normalization constants.The BP equations can be written explicitly as
\begin{widetext} 
\bea
{\hat{P}}_{i\to j}(0,0)&=&\frac{1}{{\mathcal C}_{i\to j}}\left[(1-p)\prod_{\ell\in N(i)\setminus j}[\hat{P}_{\ell\to i}(0,0)+\hat{P}_{\ell \to i}(1,0)]+p\prod_{\ell\in N(i)\setminus j}\hat{P}_{\ell \to i}(0,0)\right] \nonumber \\
\hat{P}_{i\to j}(0,1)&=&\frac{1}{{\mathcal C}_{i\to j}}\left[(1-p)\prod_{\ell\in N(i)\setminus j}[\hat{P}_{\ell\to i}(0,0)+\hat{P}_{\ell\to i}(1,0)]+pe^{-\omega}\prod_{\ell\in N(i)\setminus j}\hat{P}_{\ell\to i}(0,1)\right]\nonumber \\
\hat{P}_{i\to j}(1,1)&=&\frac{1}{{\mathcal C}_{i\to j}}pe^{-\omega}\left[\prod_{\ell\in N(i)\setminus j}[\hat{P}_{\ell\to i}(0,1)+\hat{P}_{\ell\to i}(1,1)]-\prod_{\ell\in N(i)\setminus j}\hat{P}_{\ell\to i}(0,1)\right]\nonumber \\
\hat{P}_{i\to j}(1,0)&=&\frac{1}{{\mathcal C}_{i\to j}}pe^{-\omega}\left\{\sum_{\ell\in N(i)}\hat{P}_{\ell\to i}(1,0)\prod_{\ell'\in N(i)\setminus j,\ell}\hat{P}_{\ell'\to i}(0,1)\right.\nonumber \\
&&\hspace{-5mm}\left.+\prod_{\ell\in N(i)\setminus j}\left[\hat{P}_{\ell\to i}(0,1)+P_{\ell \to i}(1,1)\right]-\prod_{\ell\in N(i)\setminus j}\hat{P}_{\ell\to i}(0,1)-\sum_{\ell\in N(i)}\hat{P}_{\ell\to i}(1,1)\prod_{\ell'\in N(i)\setminus j}\hat{P}_{\ell\to i}(0,1)\right\},
\label{BPe}
\eea
\end{widetext}
if the degree $k_i$ of node $i$ is greater than one (i.e. $k_i>1)$, whereas if the degree of node $i$ is one ($k_i=1$),  the messages are given by 
$\hat{P}_{i\to j}(0,0)=\hat{P}_{i\to j}(0,1)=1/2$ and  $\hat{P}_{i\to j}(1,0)=\hat{P}_{i\to j}(1,1)=0$.

By  solving this set of recursive equations on a given single network realization, using Eqs. $(\ref{uno})$, $(\ref{marginals})$ and $(\ref{BPe})$ it is therefore possible to determine the distribution $P(\bm{\sigma})$ in the Bethe approximation as long as the network is locally tree-like.

\subsubsection{Free energy}
The free energy of the problem can be found by minimizing the Gibbs free energy $F$  given by 
\bea
\omega F=\sum_{\bm{\sigma}}P(\bm{\sigma})\ln\left(\frac{P(\bm{\sigma})}{\psi(\bm{\sigma})}\right)
\eea
 where $\psi(\bm{\sigma})$  indicates the constraints
\bea
\psi(\bm{\sigma})=\prod_{i=1}^N\psi_i(\bm{\sigma}_i)
\eea
Indeed the  Gibbs free energy $F$ is minimal when calculated over  the probability distribution $P(\bm{\sigma})$ given by Eq. $(\ref{GM})$ when $\omega F=-\ln Z$.
By considering the Bethe approximation for the distribution $P(\bm{\sigma})$ Eq. $(\ref{uno})$, it is straightforward to see that the free energy can be expressed as 
\begin{equation}\label{betheF}
\omega F=\sum_{<i,j>}\log ({\mathcal C}_{ij}) - \sum_{i=1}^N \log({\mathcal C}_i)
\end{equation}
where the  constants ${\mathcal C}_{ij},{\mathcal C}_{i}$ can be found directly in terms of the messages $\hat{P}_{i\to j}(a,b)$, with $a,b\in \{0,1\}$. Indeed we have
\begin{widetext}
\bea
{\mathcal C}_{ij}&=&[\hat{P}_{i\to j}(0,0) \hat{P}_{j\to i}(0,0)+\hat{P}_{i \to j}(0,1)\hat{P}_{j\to i}(1,0)+\hat{P}_{j \to i }(1,0)\hat{P}_{j\to i}(0,1)+\hat{P}_{i \to j}(1,1) \hat{P}_{j\to i}(1,1)],\nonumber \\
{\mathcal C}_i&=&(1-p)\prod_{\ell\in N(i)}[\hat{P}_{\ell \to i}(0,0)+\hat{P}_{\ell\to i}(1,0)]+p\prod_{\ell \in N(i)}\hat{P}_{\ell \to i}(0.0)+pe^{-\omega}\left\{\sum_{\ell\in N(i)}\hat{P}_{\ell\to i}(1,0)\prod_{\ell'\in N(i)\setminus \ell}\hat{P}_{\ell'\to i}(0,1)+\right.\nonumber \\
&&\left.+\prod_{\ell\in N(i)}\left[\hat{P}_{\ell\to i}(0,1)+\hat{P}_{\ell \to i}(1,1)\right]-\prod_{\ell\in N(i)}\hat{P}_{\ell\to i}(0,1)-\sum_{\ell\in N(i)}\hat{P}_{\ell\to i}(1,1)\prod_{\ell'\in N(i)}\hat{P}_{\ell\to i}(0,1)\right\}.
\eea
\end{widetext}

\subsubsection{Energy and Specific Heat}

The role of the energy is played by the average size of the giant component $R$ given  by 
\bea
R&=&\sum_{\bm{\sigma}}{\mathcal R}P(\bm{\sigma})=-\frac{\partial \ln Z}{\partial \omega}.
\eea
By solving the BP equations and calculating $R$ it  is possible to observe that the system undergoes a phase transition from a non percolating phase where $R=0$ to a percolating phase where $R>0$. The set of critical points in which the transition occur is indicated by the values $(\omega_c,p_c)$  of the parameters $\omega$ and $p$.

The  specific heat $C$ is naturally defined as 
\bea
\frac{C}{\omega^2}&=&-\frac{\partial R}{\partial \omega}. \nonumber \\
\eea
where this quantity has the explicit interpretation as the variance in the   size of  giant component, i.e.
\bea
\frac{C}{\omega^2}=\left(\sum_{\bm{\sigma}}{\mathcal R}^2P(\bm{\sigma})\right)-\left(\sum_{\bm{\sigma}}{\mathcal R}P(\bm{\sigma})\right)^2.\nonumber
\eea
Both $R$ and $C/\omega^2$ can be derived from the message passing algorithm. Indeed we have 
\bea
R&=&\sum_i r_i,\label{R} \\
\frac{C}{\omega^2}&=&\sum_{i=1}^N r_i\left(1-r_i\right)
\label{C}
\eea
where 
\bea
r_i&=&\sum_{\bm{\sigma}}\rho_i P(\bm{\sigma})
\eea
indicating the probability that node $i$ is in the giant component
 is given by 
 \bea
 r_i=\frac{z_i}{{\mathcal C}_i}
 \eea
 with
\bea
&&\hspace{-9mm}z_i=pe^{-\omega}\left\{\prod_{\ell\in N(i)}\left[\hat{P}_{\ell\to i}(0,1)+\hat{P}_{\ell \to i}(1,1)\right]-\prod_{\ell\in N(i)}\hat{P}_{\ell\to i}(0,1)\right.\nonumber \\
&&\hspace{-9mm}\left.+\sum_{\ell\in N(i)}\left[\hat{P}_{\ell\to i}(1,0)-\hat{P}_{\ell\to i}(1,1)\right]\prod_{\ell'\in N(i)}\hat{P}_{\ell\to i}(0,1)\right\}.
\eea
Note that the quantity $C/\omega^2$ given by Eq. $(\ref{C})$ can be also interpreted as the fraction of nodes that given two random realizations of the initial damage are found in the giant component in one realization but not in the other. This quantity has been recently proposed \cite{Fluct1} to study the fluctuations of the giant component. Here we show that this quantity  can be naturally  interpreted as the variance of the giant component, and it is  related to the specific heat of percolation $C$.

\subsubsection{Entropy}
The entropy $S$ of the distribution is given by 
\bea
S=-\sum_{\bm{\sigma}}P(\bm{\sigma})\ln P(\bm{\sigma}),
\eea
where $P(\bm{\sigma})$ is given by the Gibbs measure $(\ref{GM})$.
From the expression of the Gibbs measure $P(\bm{\sigma})$  it   follows that  the entropy is related to the free energy by the  equation
\bea
S=\omega R+H-\omega F,
\eea
where 
\bea
H=\sum_{i=1}^N  H_i,\eea
and 
\bea
H_i=-\sum_{\bm{\sigma}_i}{\mathcal P}_{i}(\bm{\sigma}_i)\ln\left[\psi_i(\bm{\sigma}_i)\right].
\eea
The quantity $H_i$ can be expressed explicitely as a function of the messages as 
\bea
H_i&=&-\frac{(1-p)\prod_{\ell\in N(i)}[\hat{P}_{\ell \to i}(0,0)+\hat{P}_{\ell\to i}(1,0)]}{\mathcal C_i}\ln(1-p)\nonumber \\
&&\hspace{-15mm}-\left[1-\frac{(1-p)\prod_{\ell\in N(i)}[\hat{P}_{\ell \to i}(0,0)+\hat{P}_{\ell\to i}(1,0)]}{\mathcal C_i} \right]\ln p
\eea
\subsubsection{The typical scenario ($\omega=0$)}

The BP equations corresponding to $\omega=0$ reduce to the 
 the well known equations for the percolation transition characterizing the typical scenario.
In fact the BP  equations $(\ref{BPe})$ have the solution 
\bea
\hat{P}_{i\to j}(0,0)&=&\hat{P}_{i \to j} (0,1),\nonumber \\
\hat{P}_{i\to j}(1,1)&=&\hat{P}_{i\to j}(1,0).
\label{Ro0}
\eea 
As a function of $p$ we observe a phase transition  between a non-percolating phase with $R=0$, where the solution  is 
\bea
\hat{P}_{i\to j}(0,1)&=&\hat{P}_{i\to j}(0,0)=1/2,\nonumber \\
\hat{P}_{i\to j}(1,1)&=&\hat{P}_{i\to j}(1,0)=0,
\label{R0}
\eea 
and a percolating phase with $R>0$ where the solution of the BP equation is always of the type given by Eqs. $(\ref{Ro0})$ but departs from Eqs. $(\ref{R0})$.
By inserting the general  solution Eq.$(\ref{Ro0})$ in the BP equations, and 
adopting the variables 
\bea
\hat{\sigma}_{i\to j}=\hat{P}_{i\to j}(1,1)+\hat{P}_{i\to j}(1,0),
\eea
we recover the well known message passing equations  for the typical scenario of the percolation transition \cite{Weigt,Lenka}
\bea
\hat{\sigma}_{i\to j}=p\left(1-\prod_{\ell\in N(i)\j}(1-\hat{\sigma}_{\ell\to i})\right).
\eea
In this case the probability $r_i$ that a node belongs to the giant component reads
\bea
\hat{\rho}_i=p\left(1-\prod_{\ell\in N(i)}(1-\hat{\sigma}_{\ell\to i})\right).
\eea 
The thermodynamic quantities are given by 
\bea
R&=&\sum_{i=1}^N \hat{\rho}_i,\nonumber \\
\frac{C}{\omega^2}&=&\sum_{i=1}^N \hat{\rho}_i(1-\hat{\rho}_i),\nonumber \\
F&=&0,\nonumber \\
S&=&-(1-p)\ln(1-p)-p\ln p.
\eea

\section{Large deviation theory of percolation on random networks}
\label{secRand}
\subsection{Equations on random network ensemble}
The  BP equations can be studied  over a random network with degree distribution $P(k)$.
To this end we write the equations for the average messages
 \bea
 \hat{y}_{\tau}=\overline{\hat{P}_{i\to j}(\tau)}
 \eea  
 where $\tau=(a,b)$ with $a,b=0,1$  and where  $\overline{\ldots}$ indicates the average over the an ensemble of random networks with degree distribution $P(k)$.
 Since the variables $(y_{00},y_{01},y_{11},y_{10})$ are not independent but are related by the identity $$\hat{y}_{10}=1-y_{00}-y_{01}-y_{11},$$ the equations for the three independent variables $(y_{00},y_{01},y_{11})$ read,

\bea
\hat{y}_{00}&=&\sum_k \frac{k}{\Avg{k}}P(k)\frac{\left[(1-p)\left(1-\hat{y}_{01}-\hat{y}_{11}\right)^{k-1}+p\hat{y}_{00}^{k-1}\right]}{d_k}\nonumber \\
\hat{y}_{01}&=&\sum_k \frac{k}{\Avg{k}}P(k)\frac{\left[(1-p)\left(1-\hat{y}_{01}-\hat{y}_{11}\right)^{k-1}+pe^{-\omega}\hat{y}_{01}^{k-1}\right]}{d_k}\nonumber \\
\hat{y}_{11}&=&\sum_k \frac{k}{\Avg{k}}P(k)\frac{pe^{-\omega} [(\hat{y}_{01}+\hat{y}_{11})^{k-1}-\hat{y}_{01}^{k-1}]}{d_k}\nonumber \\
\label{ensemble2}
\eea
with $d_k$ given by  
\bea
d_k&=&2(1-p)\left(1-\hat{y}_{01}-\hat{y}_{11}\right)^{k-1}+p\hat{y}_{00}^{k-1}\nonumber \\&&+pe^{-\omega} \left\{2(\hat{y}_{01}+\hat{y}_{11})^{k-1}-\hat{y}_{01}^{k-1}\right.\nonumber \\
&&\left.+(k-1)[1-y_{00}-\hat{y}_{01}-2\hat{y}_{11}]\hat{y}_{01}^{k-2}\right\}.
\eea

The fraction  of nodes of degree $k$  that are in the giant component, $\rho_k$ is given by 
\bea
\rho_k&=&\frac{z_k}{{\mathcal C}_k},
\eea
where 
\bea
z_k&=&pe^{-\omega}\left[(\hat{y}_{01}+\hat{y}_{11})^k-\hat{y}_{01}^{k}\right.\nonumber \\ &&\left. +k(1-\hat{y}_{00}-\hat{y}_{01}-2\hat{y}_{11})\hat{y}_{01}^{k-1}\right]\nonumber \\
{\mathcal C}_k&=&(1-p)(1-\hat{y}_{01}-\hat{y}_{11})^k+p\hat{y}_{00}^{k}+z_k.
\label{zC}
\eea
The fraction of nodes in the giant component $r=R/N$ and the normalized specific heat $c=C/N$ are given in terms of $\rho_k$ as
\bea
r&=&\sum_kP(k)\rho_k, \nonumber \\
\frac{c}{\omega^2}&=&\sum_kP(k)\rho_k\left(1-\rho_k\right).
\eea
Finally the free energy density $f=F/N$ and normalized entropy $s=S/N$ are given respectively by 
\bea
\omega f(\omega)&=&\frac{\avg{k}}{2}\ln\left[\hat{y}_{00}^2+2\hat{y}_{01}(1-\hat{y}_{00}-\hat{y}_{01}-\hat{y}_{11})+\hat{y}_{11}^2\right]\nonumber \\&&-\sum_k P(k)\ln {\mathcal C}_k,\nonumber \\
s&=&-\omega f(\omega)+\omega r+\sum_k P(k) h_k
\eea
where ${\mathcal C}_k$ is given by Eq. $(\ref{zC})$ and $h_k$ is given by 
\bea
h_k&=&-(1-p)\frac{(1-\hat{y}_{01}-\hat{y}_{11})^k}{{\mathcal C}_k}\ln(1-p)\nonumber \\
&&-\left(1-\frac{(1-p)(1-\hat{y}_{01}-\hat{y}_{11})^k}{{\mathcal C}_k}\right)\ln p.
\eea

\subsection{The transition on the random ensemble}

The nature of the percolation transition can be explored by linearizing the Eqs. $(\ref{ensemble2})$ close to the solution ${\bf \hat{y}}^{\star}=(\hat{y}_{00}^{\star},\hat{y}_{01}^{\star},\hat{y}_{11}^{\star})$.
In this way we get a linear system of equations that reads,
\bea
{\bf \hat{y}}-{\bf \hat{y}}^{\star}={\bf \hat{J}}[{\bf \hat{y}}-{\bf \hat{y}^{\star}}]
\eea
where the  $3\times 3$ Jacobian matrix ${\bf \hat{J}}$ has elements
\bea
\hat{J}_{\alpha,\beta}=\left.\frac{\partial \hat{y}_{\alpha}}{\partial \hat{y}_{\beta}}\right|_{{\bf \hat{y}}={\bf \hat{y}^{\star}}}.
\eea
with $\alpha,\beta \in \{00,01,11\}$.

This system of equations becomes unstable when the  eigenvalue $\hat{\Lambda}_{\hat{J}}$ with maximum real part  satisfies  
\bea
Re[\hat{\Lambda}_{\hat{J}}]=1.
\label{transition}
\eea
Therefore this is the condition determining together with Eqs. $(\ref{ensemble2})$ the  percolation transition.

In the typical scenario, $\omega=0$ we get that this equation studied as a function of $p$ yield the well known continuous percolation transition describing the onset of the instability of the trivial solution   ${\bf \hat{y}}^{\star}=(1/2,1/2,0)$ at 
\bea
p\frac{\Avg{k(k-1)}}{\Avg{k}}=1.
\eea
In particular the $3\times 3$ Jacobian matrix ${\bf J}$ at ${\bf \hat{y}}^{\star}=(1/2,1/2,0)$ is given by 
\bea
{\bf J}=\left(\begin{array}{ccc}p\frac{\Avg{k(k-1)}}{\Avg{k}}& 0 &0\\
0&p\frac{\Avg{k(k-1)}}{\Avg{k}}&0\\
0&0&p\frac{\Avg{k(k-1)}}{\Avg{k}}\end{array}\right).
\eea

As a function of $\omega$ we have a line of critical points. These points  correspond to a  continuous phase transition whereas Eq. $(\ref{ensemble2})$ and Eq. $(\ref{transition})$ are satisfied at the trivial solution where $R=0$. On the contrary the transition is discontinuous and hybrid with a square root singularity when the system of equations including Eqs. $(\ref{ensemble2})$ and Eq. $(\ref{transition})$ is satisfied at a non trivial solution consistent with a non-zero size of the giant component $R>0$.


\section{Application to network ensemble and real networks}

\subsection{Analytical results on regular networks}
In any given network ensemble we have shown that the  proposed  theoretical framework for fixed value $\omega=0$ predicts the well known second order phase transition as a function of $p$ describing the typical percolation scenario.

\begin{figure*}[!htb]
   \includegraphics[width=1.90\columnwidth]{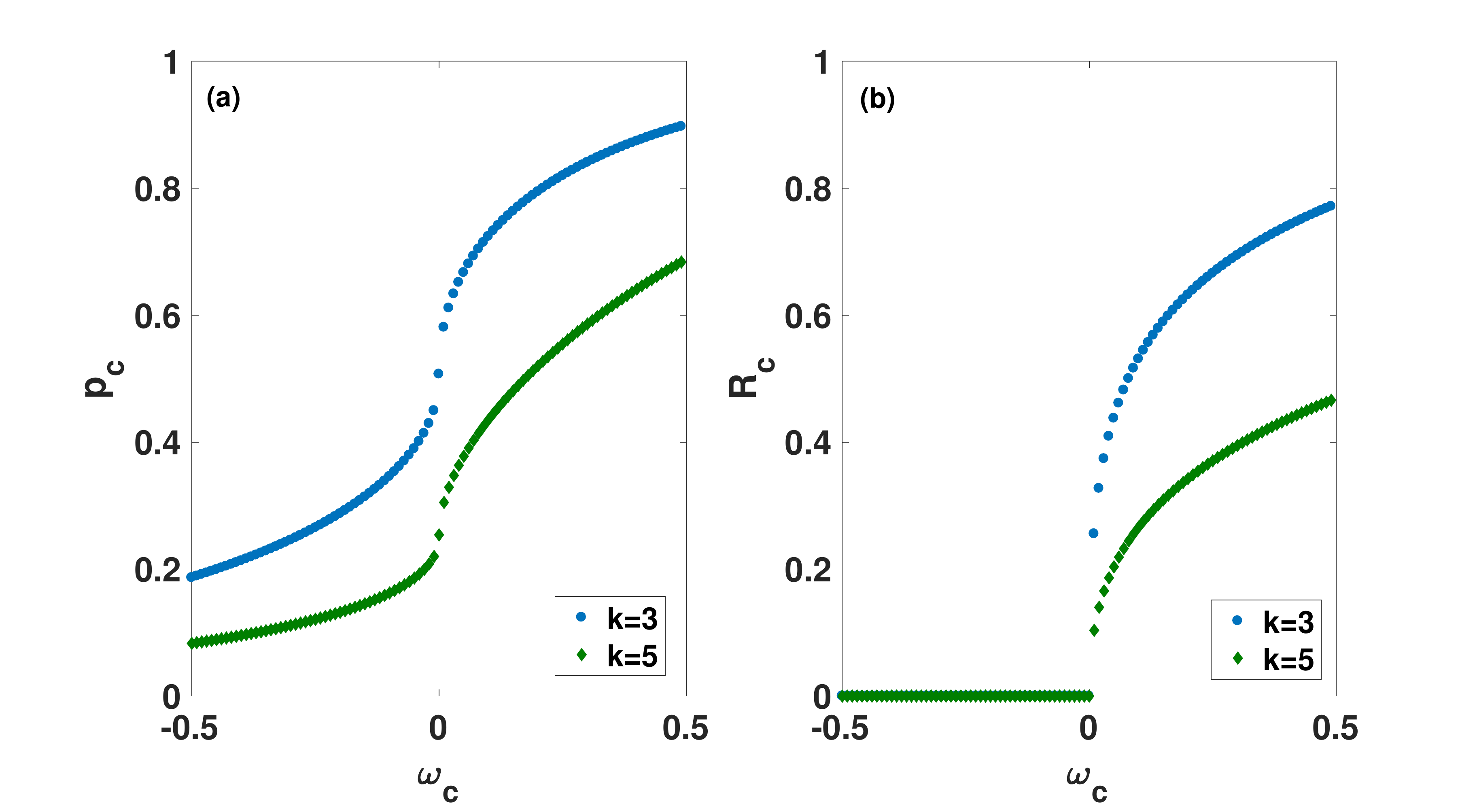}
	\caption{ The line of critical points $(\omega_c,p_c)$ (panel (a)) and value $R_c$ of the size of the giant component at the transition versus $\omega_c$ (panel (b)) are shown for a regular network with degree distribution $P(k)=\delta(k,z)$ and $z=3$ (blue circles) and $z=5$ (green diamonds). For $\omega_c>0$ the transition becomes discontinuous, i. e. $R_c>0$.}
	\label{fig:regular}
\end{figure*}

 In order to investigate the nature of the transition for $\omega \neq 0$ we have   we have numerically solved the system of  equations determining the nature of the transition (equations including Eqs. $(\ref{ensemble2})$ and Eq. $(\ref{transition})$) in the specific case of a regular network where the degree distribution is given by $P(k)=\delta(k,z)$. 
In this way  we are able to determine the phase diagram of these networks. This phase diagram reveals  that $\omega=0, p=\frac{\Avg{k}}{\Avg{k(k-1)}}, {\bf \hat{y}}^{\star}=(1/2,1/2,0)$  separates the line of continuous phase transitions from the line of discontinuous hybrid phase transitions. 
 In Figure $\ref{fig:regular}$ we  show  the   line of critical points $(\omega_c,p_c)$ for the percolation transition and the corresponding critical value  $R_c$ of the size of the giant component. The  value $R_c=0$ observed for $\omega_c \leq 0$ indicates a continuous phase transition while the values $R_c>0$ observed for $\omega_c>0$ clearly indicate discontinuous and hybrid phase transitions. 
Therefore the continuous percolation transition only characterizes the typical scenario and the configurations corresponding to $\omega<0$ but if the percolation transition is retarded ($\omega>0$) the transition becomes discontinuous.

\begin{figure*}[!htb]
    \includegraphics[width=1.90\columnwidth]{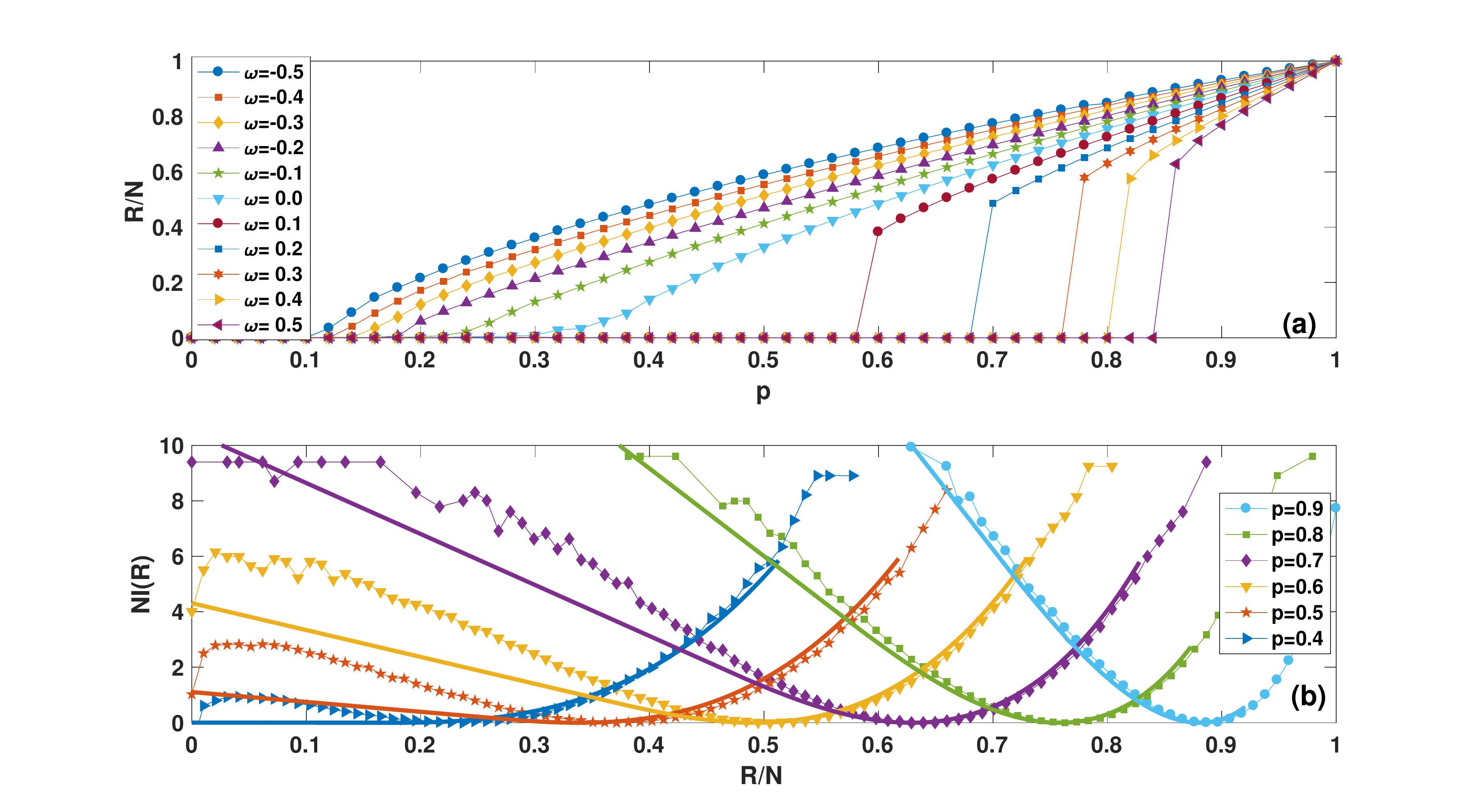}
	\caption{ The large deviation of propertie of percolation in a Poisson network with $N=100$ nodes and average degree $z=3$ are here shown to include discontinuous phase transition and non-convex rate functions. Panel (a) shows the fraction of nodes in  the giant component $R/N$ as a function of the probability that each node is not initially damaged $p$ for different values of $\omega$. For $\omega \leq 0$ a continuous percolation transition is observed, for $\omega>0$ a discontinuous percolation transition is observed. Panel (b) shows  the rate function $I(R)$ (symbols) for different values of $p$ calculated on the same network  by simulating $2\times 10^5$ realizations of the initial damage for each value of $p$. Solid lines in panel B represent the Legendre-Fenchel transform of $\omega F(\omega)/N$ which provides the convex envelop of the rate function $I(R)$. }
	\label{fig:Poisson}
\end{figure*}
\begin{figure*}[!htb]
    \includegraphics[width=1.90\columnwidth]{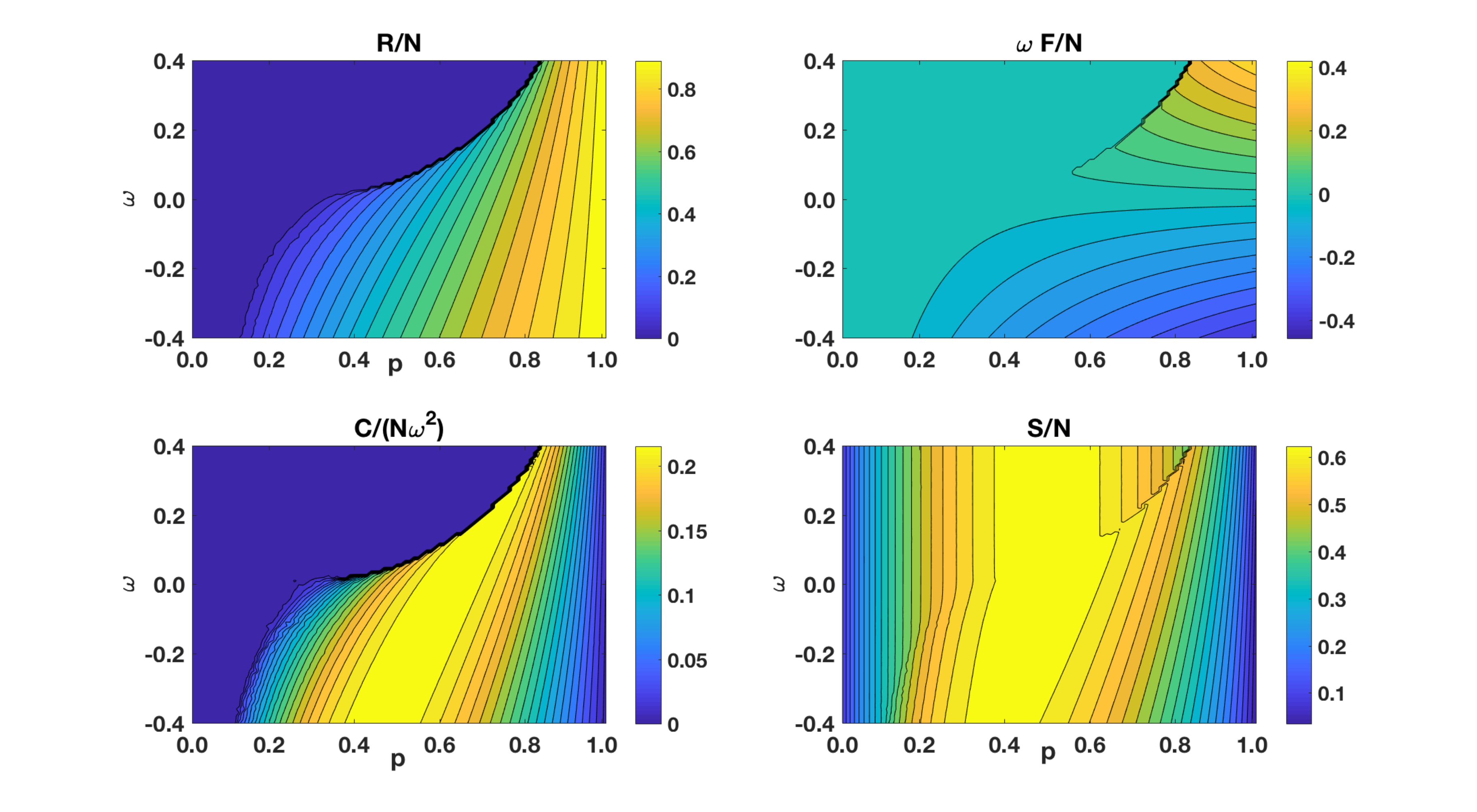}
	\caption{ The phase diagram of percolation for a Poisson network with average degree $\avg{k}=3$ and $N=100$ nodes is revealed by plotting the correspodning thermodynamic properties.  The thermodynamic quantities for a single realization of a Poisson network with $N=100$ nodes and average degree $\avg{k}=3$,    are  shown in the plane $(p,\omega)$. Here $R/N$ indicates the fraction of nodes in  the giant component, $F$ indicates the free energy, $C$ indicates the specific heat with $C/\omega^2$ given by  the variance of the size of the giant component  and $S$ indicates the entropy corresponding to a given point $(p,\omega)$ of the phase diagram.}
	\label{fig:poisson_phase_diagram}
\end{figure*}

 \begin{figure*}[!htb]
    \includegraphics[width=1.90\columnwidth]{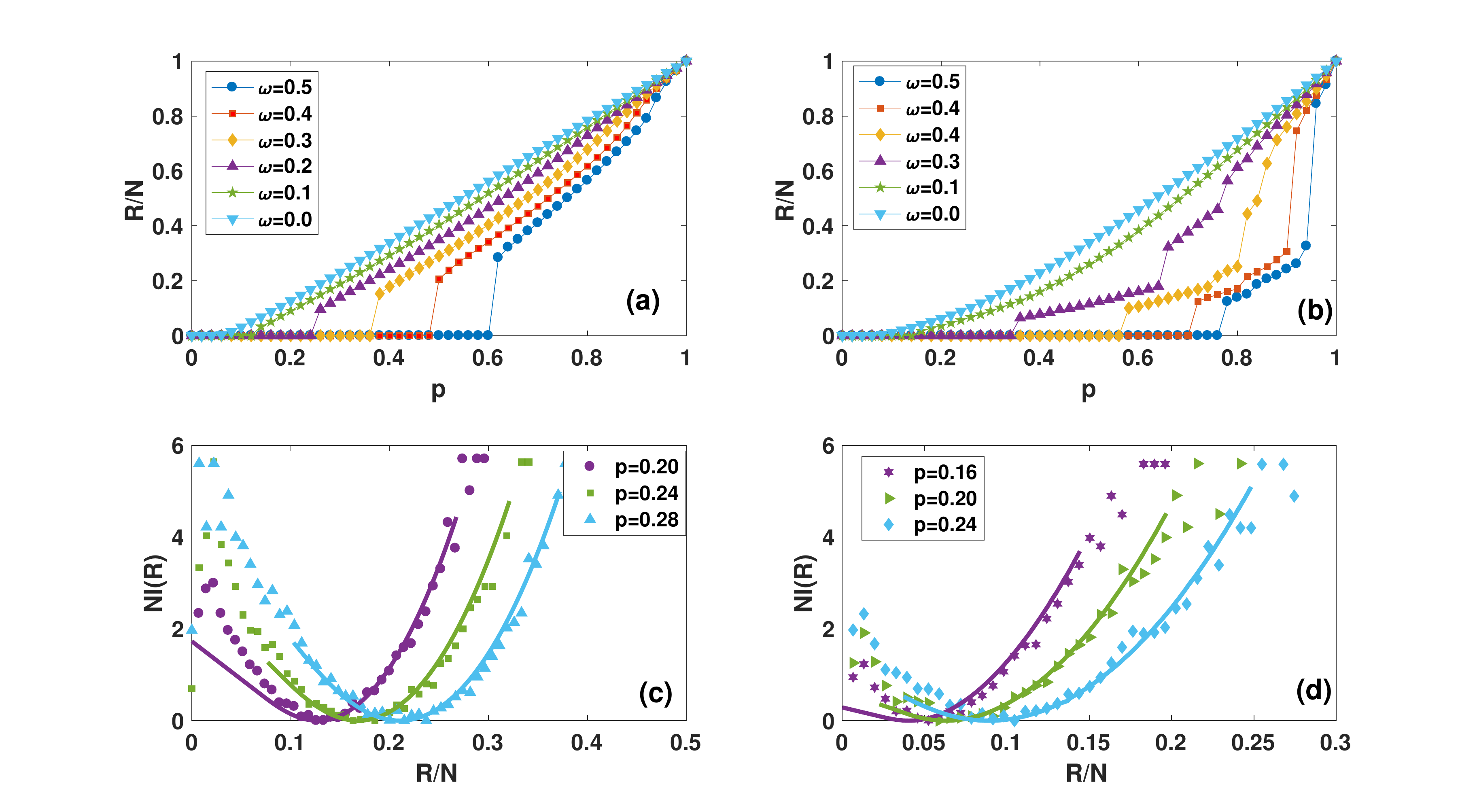}
	\caption{ The size of the giant component $R$ as a function of  $p$ for different values of  $\omega$ is shown for the Ythan Estuary (panel (a))  and  the Silwood Park (panel (b)) Foodwebs \cite{cosin}. We observe that for $\omega>0$ the percolation transition is discontinuous while for $\omega=0$ we obtain the typical scenario of percolation characterized by a continuous phase transition. The non-convex rate function $I(R)$ evaluated numerically over $2\times 10^5$ initial realization of the random damage is shown for the  Ythan Estuary (panel (c))  and  the Silwood Park (panel (d)) Foodwebs and different value of $p$ (symbols).  The  convex envelop of $I(R)$ provided by the Legendre-Fenchel transform of $\omega F(\omega)/N$ is shown with solid lines  for the same  values of $p$ (note that for improving the clarity of the figure we have omitted to plot the linear part of the convex envelop for $p=0.24,0.28$ in panel (c) and for $p=0.20,0.24$ in panel (d)).}
	\label{fig:R}
\end{figure*}

\subsection{BP results on Poisson networks and real networks}
All our numerical results of the BP algorithm on single sparse random networks and on real datasets suggest that the discontinuous phase transition for $\omega>0$ is observed generally.  Here we consider the case of a single instance of a Poisson network on which we have run the BP algorithm.
Figure  \ref{fig:Poisson}(a) shows the  predicted size of the giant component $R$  as a function of $p$ and $\omega$ for  a Poisson network with $N=100$ nodes and average degree $\avg{k}=3$.
For $\omega>0$  the giant component $R$ has a jump from a zero value $R=0$ to a non zero value $R>0$.   
Correspondingly  the rate function $I(R)$ is non-convex,  providing further evidence that the free energy $F$ is non-differentiable. In  Figure $\ref{fig:Poisson}$(b) we show the rate function $I(R)$  evaluated numerically by simulating a large number of initial damage configurations  and we compare it  to  the Legendre-Fenchel transform of the  free energy   finding very good agreement.

This investigation reveals that the observed discontinuity in the percolation problem is caused by the fact that  the rate function $I(R)$ is not convex and has a local minimum for  $R=0$ also when the expected typical size of the giant component takes positive values $\hat{R}>0$.  Therefore the rare configuration of the damage include configurations that are damaging a finite network much more than expected typically. Moreover the observed discontinuity is an indication that these configurations of the initial damage  are actually more frequent than what it might have be expected for a convex rate  function.

Additionally the   BP algorithm allows us to  characterize the entire phase diagram  of percolation using the  thermodynamics quantities $R,F,C,S$ (see Figure $\ref{fig:poisson_phase_diagram}$) fully determining the statistical mechanics properties of the percolation transition.

Finally our theoretical approach can also be used   to characterize the robustness   of real datasets against rare configuration of the random damage. In Figure $\ref{fig:R}$ we consider  two real food webs: the Ythan Estuary  (with $N=135$ nodes) and the Silwood Park  (with $N=154$ nodes) Foodwebs \cite{cosin} and we show numerical evidence for discontinuous phase transition and non-convexity of the rate function $I(R)$.
\section{Conclusions}
In conclusion  we have developed a large deviation theory for percolation on sparse networks. 
We show evidence that percolation theory, when extended to treat also the response to  rare configurations of the initial damage,  includes both continuous and discontinuous phase transitions. This result sheds light on the hidden fragility of networks and their risk of a sudden collapse and could be especially useful for understanding mechanisms to avoid the catastrophic dismantling  of real networks. The present large deviation study of percolation considers exclusively  node percolation on single networks. However the outlined methodology could be in the future extended to study the fluctuations of  generalized percolation phase transitions such as  percolation in interdependent  multilayer networks where also the typical scenario is characterized by a discontinuous phase transition.

\end{document}